\documentclass{new_tlp}
\ifx\synctex\undefined\else\synctex=1\fi
\pdfoutput=1

\usepackage{color}
\usepackage[fleqn]{amsmath} 
\usepackage{algorithm}
\usepackage[noend]{algpseudocode}
\usepackage{wrapfig}
\usepackage{tikz}
\usetikzlibrary{arrows,snakes,backgrounds,shapes,positioning,calc,intersections}
\newcommand\irregularcircle[2]{
  let \n1 = {(#1)+rand*(#2)} in
  +(0:\n1)
  \foreach \a in {10,20,...,350}{
    let \n1 = {(#1)+rand*(#2)} in
    -- +(\a:\n1)
  } -- cycle
}
\usepackage{amssymb}
\usepackage[belowskip=-10pt,aboveskip=5pt]{caption} 
\usepackage{float}
\usepackage{hyperref}

\newtheorem{theorem}{Theorem}[section]
\newtheorem{definition}{Definition}

\newtheorem{lemma}[theorem]{Lemma}
\newtheorem{proposition}[theorem]{Proposition}

\renewcommand{\phi}{\varphi}

\newcommand{\wprahft}{\textsc{WP-Rahft}}
\newcommand{\pihorn}{\textsc{PI-Horn}}
\newcommand{\trueit}{\mathit{true}}
\newcommand{\falseit}{\mathit{false}}
\newcommand{\tr}{{\sf tr}}
\newcommand{\trA}[1][A]{\ensuremath{\mathsf{tr}_{#1}}}

\newcommand{\id}[1]{\mathit{#1}}

\newcommand{\constr}{\mathsf{constr}}
\newcommand{\safe}{{\sf exit0}}
\newcommand{\unsafe}{{\sf error}}
\newcommand{\init}{{\sf init}}
\newcommand{\theory}{\mathbb{T}}
\newcommand{\trseq}{{\textsf{tr-seq}}}
\newcommand{\trseqA}[1][A]{\ensuremath{\textsf{tr-seq}_{#1}}}
\newcommand{\suffpre}[1]{{\ensuremath{\textsf{suf}}_{\textsf{#1}}}}
\newcommand{\necpre}[1]{{\ensuremath{\textsf{nec}}_{\textsf{#1}}}}

\newcommand{\tuple}[1]{\langle #1 \rangle} 
\newcommand{\tuplevar}[1]{\mathbf{#1}}

\newcommand{\pcode}[2][\codesize]{
  \fbox{
    \begin{minipage}{0.45\linewidth}
    #1
    \begin{tabbing}
    xx \= xx \= xx \= xx \= xx \= xx \= xx \= xx \= xx \= xx \= \kill
    #2
    \end{tabbing}
    \end{minipage}
  }
}

\title[Precondition Inference]{Transformation-Enabled Precondition Inference}
\author[B. Kafle et al.]
       {BISHOKSAN KAFLE \\
        IMDEA Software Institute, Madrid, Spain \\
        \email{bishoksan.kafle@imdea.org}
   \and GRAEME GANGE, PETER J. STUCKEY \\
        Faculty of IT, Monash University, Clayton Vic.\ 3800, Australia \\
        \email{\{graeme.gange,peter.stuckey\}@monash.edu}
   \and PETER SCHACHTE, HARALD S{\O}NDERGAARD \\
        School of Computing and Information Systems \\
        The University of Melbourne, Vic.\ 3010, Australia \\
        \email{\{schachte,harald\}@unimelb.edu.au}
       }
\date{}

\begin{document}
\maketitle

\begin{abstract}
Precondition inference is a non-trivial problem with important applications 
in program analysis and verification. We present a novel iterative method for 
automatically deriving preconditions for the safety and unsafety of programs. 
Each iteration maintains over-approximations of the set of \emph{safe} and 
\emph{unsafe} \emph{initial} states, which are used to partition the program's 
\emph{initial} states into those known to be \emph{safe}, known to be 
\emph{unsafe} and \emph{unknown}. 
We then construct revised programs with those \emph{unknown} initial states 
and iterate the procedure until the approximations are disjoint or some 
termination criteria are met. An experimental evaluation of the method on 
a set of software verification benchmarks shows that it can infer precise 
preconditions (sometimes optimal) that are not possible using previous methods. It is "under consideration for acceptance in TPLP".
\end{abstract}

\section{Introduction}
\label{sec:intro}

Precondition analysis infers input conditions that establish runtime properties 
of interest (for example, a \emph{sufficient precondition} for safety is a set 
of initial states, each of which is guaranteed to be safe with respect to 
given safety properties).
Applications include program verification, symbolic execution, debugging, 
and program comprehension.
Derivation of exact preconditions (excluding no good runs and including no 
bad runs) is undecidable, so the aim is to derive preconditions that are 
general enough to be useful in practice. 
We approach the problem by iteratively refining over-approximations of 
safe and unsafe states. 
For this, constrained Horn clauses (CHCs) are convenient, 
as they can model imperative programs and assertions in a uniform way 
\cite{Peralta-Gallagher-Saglam-SAS98,DBLP:conf/pldi/GrebenshchikovLPR12,DBLP:conf/cav/GurfinkelKKN15,DBLP:journals/scp/AngelisFPP17}. 

\begin{figure}[t]
  \begin{center}
  \begin{tabular}{ll}
    \pcode[\small]{
   \textbf{int} main(\textbf{int}~ a, \textbf{int}~ b) \{ \\
   ~~\textbf{while} ($a \geq 1 $) \{  \\
   ~~~~$a=a-1;$ 
      	 $b=b-1;$ \\    
   ~~\}~~ \\
   ~~\textbf{assert} ($b \geq 0$); \\
   \} 
      }    &
    \pcode[\small]{
$\mathtt{c_1.~ init(A,B)}.$ \\ 
$\mathtt{c_2.~ wh(A,B) \leftarrow init(A,B).}$\\
$\mathtt{c_3.~ wh(A,B) \leftarrow A_0\geq1, A=A_0-1,}$ \\
$\mathtt{~~~~~~~~~~~~~~~~~~~~ B=B_0-1, wh(A_0,B_0).}$\\
$\mathtt{c_4.~ error\leftarrow A<1, B<0, wh(A,B).}$ \\
$\mathtt{c_5.~ exit0\leftarrow A<1, B\geq0, wh(A,B).}$
    } 
  \end{tabular}
  \end{center}
  \caption{Running example: (left) original program, (right) translation to
   CHCs}
 \label{ex:precond}
\end{figure}

Consider the program in Fig.~\ref{ex:precond}.
The left box shows a fragment in C, the right box shows its 
CHC representation, encoding reachable states.
C variables are represented by logical variables (capital letters). 
The clause $c_1$ specifies the initial states of the program via the 
predicate \emph{init} which is always reachable. 
Similarly, $c_2$ and $c_3$ encode the reachability of the \emph{while} 
loop via the predicate \emph{wh}. 
Clause $c_2$ states that the loop is reachable if \emph{init} is 
reachable while $c_3$ states that the loop is (re-)reachable from the 
end of its own body (recursive case). 
The last two clauses represent the properties of the program. 
Clause $c_4$ states that an ``unsafe or error'' state is reached if 
$\mathtt{B<0}$ upon loop exit (encoded by $\mathtt{error}$), 
and the clause $c_5$ states that the program terminates gracefully or 
reaches a safe state if $\mathtt{B \geq 0}$ (encoded by $\mathtt{exit0}$). 
The program is \emph{unsafe} if it reaches $\mathtt{error}$ for some 
input and is \emph{safe} if none of the input reaches $\mathtt{error}$. 
(The semantics of \emph{assert(c)} is 
\emph{if(c)} \texttt{SKIP} \emph{else} \texttt{ERROR}.)

Clearly the program terminates.
Its assertion is violated if the initial conditions on \verb~a~ and \verb~b~ 
entail the disjunction 
$\mathtt{(b<0 \wedge a \leq 0) \vee (a\geq 1 \wedge a> b)}$.
It terminates gracefully if 
$\mathtt{(b\ge0 \wedge a\leq 0) \vee (a\geq 1 \wedge b\geq a)}$. 
Automatic derivation of these preconditions is challenging for at least 
three reasons:
\begin{enumerate}
\item[(i)]
The desired result is a disjunction of linear constraints---to reach it,
we need the ability to express disjunctive information.
\item[(ii)]
Invoking an abstract interpreter using forward analysis on the original 
program derives $a \geq 1$ as invariant for predicate $\mathtt{wh}$, 
while an abstract interpreter working backward from a goal such as
$\mathtt{exit0}$ derives $b \geq 1$ as invariant for the same predicate.
That is, without a more sophisticated approach, we lose critical information 
about \verb~a~ and \verb~b~ in backward and forward analysis, respectively. 
\item[(iii)]
We need to reason simultaneously about the safe and unsafe states; 
one type of information cannot simply be obtained by complementing the other. 
For example, the formula 
$\mathtt{(b \ge 0 \wedge a \leq 0) \vee (a \geq 1 \wedge b \geq a)}$ 
that implies safe termination cannot be obtained by negating the formula 
$\mathtt{(b<0 \wedge a \leq 0) \vee (a\geq 1 \wedge a> b)}$ 
that implies violation of the assertion. 
Previous approaches \cite{DBLP:conf/lopstr/HoweKL04,DBLP:conf/vmcai/Moy08,DBLP:journals/entcs/Mine12,DBLP:conf/sas/BakhirkinBP14,kafle-iclp18} 
fail to infer the desired preconditions, either because they can only
infer conjunctive information, or because they rely on an abstract operation
of complementation that comes with some loss of precision.
\end{enumerate}
The method we present addresses the challenges as follows.
Challenge (i) is addressed via partial evaluation that creates a finite 
number of versions of each predicate, which is essential for deriving 
disjunctive invariants. 
Many loops in typical program patterns, such as our example, 
require disjunctive invariants to be established.
While there are techniques that are capable of inferring precise invariants
\cite{SankaranarayananSM04,GulwaniJK09,GuptaR09,BeyerHMR07},
we argue that they are significantly more complicated, less efficient, 
and less widely-used than standard abstract interpretation-based techniques 
for generating conjunctive invariants \cite{Karr76,CousotH78,Mine06}. 
In this article we use the standard invariant generation tools in conjunction 
with program transformations with the aim of raising their precision level
to that of disjunctive invariant generation tools.
Challenge (ii) is addressed via forward and backward abstract interpretation, 
together with constraint specialisation. 
Using this, one can infer $a \geq 1, b \geq a$ as invariant for predicate 
$\mathtt{wh}$, as described in Section~\ref{sec:specialisation}.
Challenge (iii) is addressed by concurrently maintaining and refining 
approximations of both safe and unsafe states, until the approximations 
are disjoint or some termination criterion is met.  

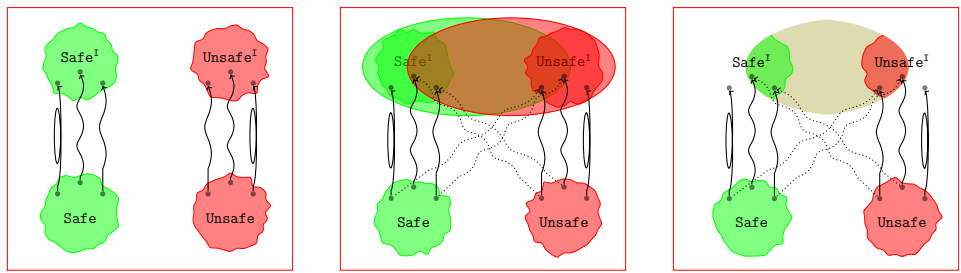
\begin{figure}[t]
\centering
\begin{minipage}{.32\textwidth}
\tikz \node[scale=0.5, inner sep=0]{
\begin{tikzpicture}
[align=center,node distance=4cm and 6cm]
  \tikzstyle{ellipse} = [ellipse, minimum height=8em, minimum width=5em, draw,thick]
\tikzset{dot/.style={circle,fill=#1,inner sep=0,minimum size=4pt}}
  \tikzstyle{block} = [rectangle, rounded corners, minimum height=1em, minimum width=6em, draw]
 
\coordinate (safeC) at (-3,-3);
\coordinate[right of=safeC]  (unsafeC) ;
\coordinate[above of=safeC] (isafeC) ;
\coordinate[above of=unsafeC] (iunsafeC) ;
  
  \matrix {

  \begin{scope}[fill opacity=0.5]
  \draw[green,fill,rounded corners=.5mm,name=safe]  (safeC) \irregularcircle{1cm}{1mm} {} ;
  \node[below,black,opacity=1.0] at (safeC.south){\large$\mathtt{Safe}$};
  \draw[red,fill,rounded corners=.5mm,name=unsafe]  (unsafeC) \irregularcircle{1cm}{1mm} {};
  \node[below,black,opacity=1.0] at (unsafeC.south){\large$\mathtt{Unsafe}$};
  \draw[green,fill,rounded corners=.5mm,name=isafe]  (isafeC) \irregularcircle{1cm}{1mm} {} ;
  \node[above,black,opacity=1.0] at (isafeC.south){\large$\mathtt{Safe^I}$};
  \draw[red,fill,rounded corners=.5mm,name=iunsafe]  (iunsafeC) \irregularcircle{1cm}{1mm} {};
  \node[above,black,opacity=1.0] at (iunsafeC.south){\large$\mathtt{Unsafe^I}$};
    
  \node[dot] (s1) at (-3,-2.2) {};
  \node[dot] (s2) at (-3.6,-2.5){};
  \node[dot] (s3) at (-2.4,-2.5){};
  
  \node[dot] (is1) [above of=s1,yshift=-30pt]{};
  \node[dot] (is2) [above of=s2,yshift=-30pt]{};
  \node[dot] (is3) [above of=s3,yshift=-30pt]{};
  
  \node[dot] (iu1) [right of=is1]{};
  \node[dot] (iu2) [right of=is2]{};
  \node[dot] (iu3) [right of=is3]{};
  
  \node[dot] (u1) [right of=s1]{};
  \node[dot] (u2) [right of=s2]{};
  \node[dot] (u3) [right of=s3]{};
  \draw [<-,thick,snake=coil,segment aspect=10,segment length=17pt] (is2) -- (s2)  {} ;
  \draw [<-,thick,snake=coil,segment length=30pt] (is1) -- (s1)  {} ;
  \draw [<-,thick,snake=coil,segment length=40pt] (is3) -- (s3)  {} ;
       
  \draw [<-,thick,snake=coil,segment length=30pt] (iu1) -- (u1) node[midway,left] {} ;
  \draw [<-,thick,snake=coil,segment length=40pt] (iu2) -- (u2) node[midway,left] {} ;
  \draw [<-,thick,snake=coil,segment aspect=10,segment length=17pt] (iu3) -- (u3) 
	node[midway,left] {} ; 
    \end{scope}
\\};

\begin{pgfonlayer}{background}
\draw [draw=red]
([xshift=-27pt,yshift=-15pt]safeC.south)

rectangle ([xshift=75pt,yshift=70pt]iunsafeC.north);
\end{pgfonlayer}
\end{tikzpicture}
};
\end{minipage}
\begin{minipage}{.32\textwidth}
\tikz \node[scale=0.5,inner sep=0]{
\begin{tikzpicture}
[align=center,node distance=4cm and 6cm]
  \tikzstyle{sharp} = [ellipse,minimum height=8em,minimum width=17em,draw,thick,fill]
\tikzset{dot/.style={circle,fill=#1,inner sep=0,minimum size=4pt}}
  \tikzstyle{block} = [rectangle,rounded corners,minimum height=1em,minimum width=6em,draw]
 
\coordinate (safeC) at (-3,-3);
\coordinate[right of=safeC]  (unsafeC) ;
\coordinate[above of=safeC] (isafeC) ;
\coordinate[above of=unsafeC] (iunsafeC) ;
  
  \matrix {
  
  \begin{scope}[fill opacity=0.5]
  \draw[green,fill,rounded corners=.5mm,name=safe]  (safeC) \irregularcircle{1cm}{1mm} {} ;
  \node[below,black,opacity=1.0] at (safeC.south){\large$\mathtt{Safe}$};
  \draw[red,fill,rounded corners=.5mm,name=unsafe]  (unsafeC) \irregularcircle{1cm}{1mm} {};
  \node[below,black,opacity=1.0] at (unsafeC.south){\large$\mathtt{Unsafe}$};
  \draw[green,fill,rounded corners=.5mm,name=isafe]  (isafeC) \irregularcircle{1cm}{1mm} {} ;
  \node[above,black,opacity=1.0] at (isafeC.south){\large$\mathtt{Safe^I}$};
  \node[sharp,draw,green] (safesharp) at ([xshift=40pt]isafeC){};
  \draw[red,fill,rounded corners=.5mm,name=iunsafe]  (iunsafeC) \irregularcircle{1cm}{1mm} {};
  \node[above,black,opacity=1.0] at (iunsafeC){\large$\mathtt{Unsafe^I}$};
  \node[sharp,draw,red] (unsafesharp) at ([xshift=-40pt]iunsafeC){};
    
  \node[dot] (s1) at (-3,-2.2) {};
  \node[dot] (s2) at (-3.6,-2.5){};
  \node[dot] (s3) at (-2.4,-2.5){};
  
  \node[dot] (is1) [above of=s1,yshift=-30pt]{};
  \node[dot] (is2) [above of=s2,yshift=-30pt]{};
  \node[dot] (is3) [above of=s3,yshift=-30pt]{};
  
  \node[dot] (iu1) [right of=is1]{};
  \node[dot] (iu2) [right of=is2]{};
  \node[dot] (iu3) [right of=is3]{};
  
  \node[dot] (u1) [right of=s1]{};
  \node[dot] (u2) [right of=s2]{};
  \node[dot] (u3) [right of=s3]{};
  \draw [<-,thick,snake=coil,segment aspect=10,segment length=17pt] (is2) -- (s2)  {} ;
  \draw [<-,thick,snake=coil,segment length=30pt] (is1) -- (s1)  {} ;
  \draw [<-,thick,snake=coil,segment length=40pt] (is3) -- (s3)  {} ;
       
  \draw [<-,thick,snake=coil,segment length=30pt] (iu1) -- (u1) node[midway,left] {} ;
  \draw [<-,thick,snake=coil,segment length=40pt] (iu2) -- (u2) node[midway,left] {} ;
  \draw [<-,thick,snake=coil,segment aspect=10,segment length=17pt] (iu3) -- (u3) node[midway,left] {} ; 
       
  \draw [<-,dotted,thick,snake=coil,segment length=30pt] (is1) -- (u2)  {} ;
  \draw [<-,dotted,thick,snake=coil,segment length=40pt] (is3) -- (u1)  {} ;
  \draw [<-,dotted,thick,snake=coil,segment length=30pt] (iu1) -- (s3)  {} ;
  \draw [<-,dotted,thick,snake=coil,segment length=40pt] (iu2) -- (s2)  {} ;
  \end{scope}
\\};

\begin{pgfonlayer}{background}
\draw [draw=red]
([xshift=-27pt,yshift=-15pt]safeC.south)

rectangle ([xshift=75pt,yshift=70pt]iunsafeC.north);
\end{pgfonlayer}
\end{tikzpicture}
};
\end{minipage}
\begin{minipage}{.32\textwidth}

\tikz \node[scale=0.5,inner sep=0]{
\begin{tikzpicture}
[align=center,node distance=4cm and 6cm]
  \tikzstyle{sharp} = [ellipse,minimum height=8em,minimum width=17em,draw,thick,fill]
\tikzset{dot/.style={circle,fill=#1,inner sep=0,minimum size=4pt}}
\tikzstyle{block} = [rectangle,rounded corners,minimum height=1em,minimum width=6em,draw]
 
\coordinate (safeC) at (-3,-3);
\coordinate[right of=safeC] (unsafeC) ;
\coordinate[above of=safeC] (isafeC) ;
\coordinate[above of=unsafeC] (iunsafeC) ;
  
  \matrix {
  
  \begin{scope}[fill opacity=0.5]
  \draw[green,fill,rounded corners=.5mm,name=safe] (safeC) \irregularcircle{1cm}{1mm} {} ;
  \node[below,black,opacity=1.0] at (safeC.south){\large$\mathtt{Safe}$};
  \draw[red,fill,rounded corners=.5mm,name=unsafe] (unsafeC) \irregularcircle{1cm}{1mm} {};
  \node[below,black,opacity=1.0] at (unsafeC.south){\large$\mathtt{Unsafe}$};
  \begin{scope}
    \path[clip] ([xshift=40pt]isafeC) ellipse [x radius=8.5em,y radius=4em];
    \path[clip] ([xshift=-40pt]iunsafeC) ellipse [x radius=8.5em,y radius=4em];
    \draw[fill,olive!60] ($(isafeC)!0.5!(iunsafeC)$) circle (20em);
    \draw[green,fill,rounded corners=.5mm,name=isafe]  (isafeC) \irregularcircle{1cm}{1mm} {} ;
    \draw[red,fill,rounded corners=.5mm,name=iunsafe]  (iunsafeC) \irregularcircle{1cm}{1mm} {};
  \end{scope}
  \node[above,black,opacity=1.0] at (isafeC.south){\large$\mathtt{Safe^I}$};
  \node[above,black,opacity=1.0] at (iunsafeC){\large$\mathtt{Unsafe^I}$};
    
  \node[dot] (s1) at (-3,-2.2) {};
  \node[dot] (s2) at (-3.6,-2.5){};
  \node[dot] (s3) at (-2.4,-2.5){};
  
  \node[dot] (is1) [above of=s1,yshift=-30pt]{};
  \node[dot] (is2) [above of=s2,yshift=-30pt]{};
  \node[dot] (is3) [above of=s3,yshift=-30pt]{};
  
  \node[dot] (iu1) [right of=is1]{};
  \node[dot] (iu2) [right of=is2]{};
  \node[dot] (iu3) [right of=is3]{};
  
  \node[dot] (u1) [right of=s1]{};
  \node[dot] (u2) [right of=s2]{};
  \node[dot] (u3) [right of=s3]{};
  \draw [<-,thick,snake=coil,segment aspect=10,segment length=17pt] (is2) -- (s2)  {} ;
  \draw [<-,thick,snake=coil,segment length=30pt] (is1) -- (s1)  {} ;
  \draw [<-,thick,snake=coil,segment length=40pt] (is3) -- (s3)  {} ;
       
  \draw [<-,thick,snake=coil,segment length=30pt] (iu1) -- (u1) node[midway,left] {} ;
  \draw [<-,thick,snake=coil,segment length=40pt] (iu2) -- (u2) node[midway,left] {} ;
  \draw [<-,thick,snake=coil,segment aspect=10,segment length=17pt] (iu3) -- (u3) node[midway,left] {} ; 
       
  \draw [<-,dotted,thick,snake=coil,segment length=30pt] (is1) -- (u2)  {} ;
  \draw [<-,dotted,thick,snake=coil,segment length=40pt] (is3) -- (u1)  {} ;
  \draw [<-,dotted,thick,snake=coil,segment length=30pt] (iu1) -- (s3)  {} ;
  \draw [<-,dotted,thick,snake=coil,segment length=40pt] (iu2) -- (s2)  {} ;
     
  \end{scope}
\\};

\begin{pgfonlayer}{background}
\draw [draw=red]
([xshift=-27pt,yshift=-15pt]safeC.south)

rectangle ([xshift=75pt,yshift=70pt]iunsafeC.north);
\end{pgfonlayer}
\end{tikzpicture}
};
\end{minipage}
\vspace{2mm}
\caption{Precondition inference: Reality (left), initial approximations (middle), one step refinement of approximations using Algorithm \ref{alg:precond} (right). Arrows represent preconditions flowing from safe or unsafe final states (bottom) back to corresponding initial states (top).
\label{fig:alg-progress}}
\end{figure}

Fig.~\ref{fig:alg-progress} sketches the idea. 
The leftmost panel reflects reality, 
showing the set of concrete \emph{safe} and \emph{unsafe} states, 
along with the corresponding sets of \emph{initial} states.
Given a program and a description of the sets of interest 
(safe and unsafe states), precondition analysis infers a set of initial 
states that lead to these sets of interest.
Or rather, it finds \emph{over-approximations} of the initial sets of 
states---so these may overlap. 
Over-approximations are shown as ellipses in the middle panel, 
coloured appropriately for safe and unsafe initial states. 
Because of approximation, there may be witness traces from the left ellipse 
to $\mathtt{Unsafe}$ (the dotted arrows) and vice versa. 
The algorithm aims to reduce the ellipses progressively to the point 
where the ellipses no longer overlap. 
A single refinement step that focuses our attention only on the 
intersection is illustrated in the rightmost panel.

Our work builds upon the transformation-guided framework of 
\citeN{kafle-iclp18} and incorporates a number of program 
transformations known from the literature, including 
\begin{itemize}
\item 
Partial evaluation \cite{Jones-Gomard-Sestoft}: PE wrt.\ a goal 
specialises a program for the given goal; preserving only those 
derivations that are relevant for deriving the goal. 
\item 
Constraint Specialisation \cite{DBLP:journals/scp/KafleG17} via forward
and backward abstract interpretation \cite{DBLP:conf/sas/BakhirkinM17}: 
This strengthens constraints in clauses by exploiting generated invariants, 
while preserving derivations of a goal. 
The effect is to prune paths that are not relevant for deriving the goal.
\item 
Trace Elimination \cite{DBLP:journals/cl/KafleG17}: 
This eliminates a set of traces from a program while preserving the 
rest of traces and serves to refine a program.
\end{itemize}
Our contribution is to combine these techniques into an iterative framework 
that can control the quality of the preconditions for both safety and unsafety. 
\citeN{kafle-iclp18} iteratively apply CHC transformations to a program wrt.\
$\unsafe$, approximating the unsafe states whose complement yields sufficient 
preconditions for the safety.  
A disadvantage of this is the \emph{blind} refinement of unsafe states 
without knowing its frontier with the safe states. 
This misses opportunities to avoid redundant computation as well as to 
guide the refinement process at an early stage.
We extend that work in a number of directions: 
\begin{itemize}
\item 
We model both the safe and unsafe program states, and refine them concurrently,
allowing us to derive preconditions for both safety and unsafety. 
In addition, we show how to derive preconditions for both safety and unsafety 
either from the original program or one obtained via a sequence of 
transformations (\S \ref{sec:specialisation}).
\item 
We present an iterative algorithm to refine approximations of these states,
each iteration focusing only on states yet to be shown safe or unsafe 
(the intersection of safe and unsafe over-approximations), 
thus reducing the search space. 
It has refined termination criteria to control precision and detect 
optimality of the preconditions.
\item 
Reasoning simultaneously about safe and unsafe states allows us to 
derive precondition for non-termination as a complement of necessary 
preconditions for safety and unsafety (\S \ref{sec:algorithm}). 
\item 
Evaluation shows that we not only infer non-trivial preconditions in 
slightly more cases but also infer optimal ones in some cases 
(\S \ref{sec:experiments}). 
\end{itemize} 

\section{Preliminaries}
\label{sec:prelim}
An \emph{atom} is a formula $p(\tuplevar{x})$ 
where $p$ is a predicate symbol and $\tuplevar{x}$ a tuple of arguments. 
A constrained Horn clause (CHC) is a first-order formula written as 
$p_0(\tuplevar{x}_0) \leftarrow 
  \phi, p_1(\tuplevar{x}_1), \ldots, p_k(\tuplevar{x}_k)$ 
following Constraint Logic Programming (CLP) standard, 
where $\phi$ is a finite conjunction of quantifier-free \emph{constraints} 
on variables $\tuplevar{x}_i$ with respect to some constraint theory 
$\theory$, $p_i(\tuplevar{x}_i)$ are atoms. 
A \emph{constrained fact} is a clause of the form 
$p_0(\tuplevar{x}_0) \leftarrow \phi$, where $\phi$ is a constraint.  
We assume the theory $\theory$ is equipped with a decision procedure 
and a projection operator, and that it is closed under negation.

The notation $\phi|_{V}$ represents the constraint formula $\phi$
projected onto variable set $V$ and $\phi \models_{\theory} \psi$ 
(or equivalently $ \models_{\theory}\phi \rightarrow \psi$) 
to represent $\phi$ entails $\psi$ over $\theory$. 
Similarly, we write $P \vdash_{\theory}A$ when an atom $A$ is derivable 
from the program $P$ wrt.\ the theory $\theory$.
 
We use CHCs to encode control flow of C-like programs.
Two special predicates, $\safe$ and $\unsafe$, encode safe and unsafe 
(error) states, respectively.
So $\safe$ indicates a normal return; 
$\unsafe$ indicates abnormal termination.
The predicate \init\ encodes the set of initial states. 
We assume users specify all states of interest by appropriate constructs 
provided by the language (e.g., \emph{assert(c), return $\tuple{n}$} of C). 
States not specified by the user 
(e.g., \emph{buffer-overflow, floating point exceptions}) 
are not taken into account while generating CHCs. 
Hence correctness of the preconditions depends on the user 
specified set of states. 
  
From here on, when talking about a program, we refer to its 
CHC representation. 

\begin{definition}[AND-tree \cite{Gallagher-Lafave-Dagstuhl}]\label{def:andtree}
\rm
An \emph{AND-tree} for a CHC program $P$ is a tree whose nodes are 
labelled as follows.
\begin{enumerate}
\item
Each non-leaf node corresponds to a (renamed) clause in $P$
of the form $A \leftarrow \phi, A_1,\ldots,A_k$ where $k>0$.
The clause is renamed so that any variables not appearing in $A$ are fresh.
The node is labelled by $(A, \phi)$.
The node has $k$ child nodes where the $i^{th}$ child corresponds to a 
clause in $P$ of the form $A_i \leftarrow \phi_i, B_i$ 
where $B_i$ may be empty and is labelled by $(A_i,\phi_i)$. 
\item
Each leaf node corresponds to a (renamed) clause in $P$ of the form 
$A \leftarrow \phi$ and is labelled as ($A, \phi$). 
\end{enumerate}
 \end{definition}
Given an AND-tree $t$, $\constr(t)$ is the conjunction of the constraints
appearing in the tree. 
The tree $t$ is \emph{feasible} if and only if $\constr(t)$ is satisfiable 
over $\theory$.

\begin{definition}[Initial clauses and nodes]\label{def:initial-node}
\rm
Let $P$ be a program with a {distinguished} predicate 
$p^{I}$ which we call the \emph{initial predicate}.
The \emph{constrained facts}
$\{(p^I(\tuplevar{x}) \leftarrow \theta) 
\mid (p^I(\tuplevar{x}) \leftarrow \theta) \in P\}$ 
are the \emph{initial clauses} of $P$.
Let $t$ be an AND-tree for $P$.
A node labelled by 
$p^I(\tuplevar{x}) \leftarrow \theta$ is an \emph{initial node} of $t$.
We extend the term ``initial predicate" and use the symbol $p^I$ to 
refer also to renamed versions of the initial predicate that arise 
during clause transformations.
\end{definition}

In Fig.~\ref{ex:precond} the initial predicate is \texttt{init} and 
the initial clause is $\mathtt{ init(A,B)}$.

\section{Program transformations and preconditions}
\label{sec:specialisation}

We now show how to find preconditions for safety and unsafety 
for a program (original or obtained via transformation).
We limit attention to sets of clauses for which every AND-tree for $\safe$ 
and $\unsafe$ (whether feasible or infeasible) has at least one initial node. 

\begin{definition}[Program with initial states $\phi$ ($P_{\phi}^I$)
and replaced states $\phi$ ($P_{\phi}^R$)]
\label{def:repis}
\rm
Let $P$ be a program and $\phi$ a constraint over $\theory$. 
Let $P_{\phi}^I$ be the 
clauses obtained from $P$ by replacing the initial clauses 
$\{(p^I(\tuplevar{x}) \leftarrow \theta_i) \mid 1 \le i \le k\}$ by 
$\{(p^I(\tuplevar{x}) \leftarrow \phi \wedge \theta_i) \mid 1 \le i \le k\}$. 
Similarly, let $P_{\phi}^R$ be the set of 
clauses obtained from $P$ by replacing the initial clauses 
$\{(p^I(\tuplevar{x}) \leftarrow \theta_i) \mid 1 \le i \le k\}$ by 
$\{(p^I(\tuplevar{x}) \leftarrow \phi) \}$.  
\vspace*{-1ex}
\end{definition}

\begin{definition}[Necessary/sufficient precondition for safety]
\label{def:nec-suff-precond}
\rm
Let $P$ be a program and $\phi$ a constraint over $\theory$. 
Then 
\begin{itemize}
\item a constraint $\phi$ is a \emph{necessary precondition} 
(NP) for the safety of $P$ 
if $P \vdash_\theory \safe$ entails $P_{\phi}^R \vdash_\theory \safe$. 
In words, $\phi$ (possibly $\trueit$) is an over-approximation of 
the set of initial states of $P$ that can reach \safe. 

\item a constraint $\psi$ is a \emph{sufficient precondition} (SP) 
for the safety of $P$ 
if $P_\psi^I \not \vdash_\theory \unsafe$. 
In words, $\psi$ (possibly $\falseit$) is an under-approximation 
of the set of initial states of $P$ that cannot reach \unsafe. 
\end{itemize}
\end{definition}
Thus an SP for safety is a constraint that 
suffices to block derivations of $\unsafe$ (given we assume 
clauses for which $p^I$ is essential for any derivation of $\unsafe$). 
In practice we would like to consider SP for safety as a constraint that
allows derivations of \safe\ \emph{and} blocks derivations of $\unsafe$.
We define NP ($\necpre{u}{(P)}$) and SP ($\suffpre{u}{(P)}$) for unsafety 
analogously.
In the following, we show how an NP and an SP can be derived from a set 
of clauses.

\begin{definition}[NP extracted from CHC program $P$] 
\label{def:necsafe} 
\rm
Let $P$ be a set of clauses encoding reachable states of a program. 
The formula 
\[
  \bigvee \{\theta \mid (p^I(\tuplevar{x}) \leftarrow \theta) \in P\}.
\]
is an NP for \emph{both} safety and unsafety. 
We refer to it as $\necpre{s}(P)$ when talking about safety and as
$\necpre{u}(P)$ when talking about unsafety.
\end{definition}
The reason why $\necpre{s}(P)$ is an NP for safety is that any feasible 
AND-tree for $P$ must use at least one initial clause of $P$ and so the 
disjunction of constraints from the initial clauses (although imprecise) 
is a sound NP for the safety, as well as the unsafety, of $P$. 
Using this definition, the NPs for the program $P$ in Fig.~\ref{ex:precond} 
are $\necpre{s}{(P)}=\necpre{u}{(P)}=\trueit$ 
(the set of initial states of the program). 
Given NPs for a program $P$, we can find \emph{sufficient preconditions} 
for the safety and unsafety as follows.

\begin{definition}[SP extracted from CHC program $P$]
Let $P$ be a set of clauses encoding reachable states of a program. 
We define 
\[
\begin{array}{lcl}
   \suffpre{s}{(P)} &=& \necpre{s}(P) \wedge \neg \necpre{u}(P)
\\ \suffpre{u}{(P)} &=& \necpre{u}(P) \wedge \neg \necpre{s}(P)
\end{array}
\]
\end{definition}
The former is a sufficient precondition for safety, the latter for unsafety,
of $P$. In the sequel, we represent necessary and sufficient preconditions 
by Greek letters $\phi$ and $\psi$, respectively, 
possibly with subscript \emph{u} for unsafety and \emph{s} for safety.

Let $\phi_c$ be $\necpre{s}(P) \wedge \necpre{u}(P)$, denoting the shared 
region between the (approximate) safe and unsafe states. 
A precondition is \emph{separating} (optimal) if $\phi_c$ is 
\emph{unsatisfiable} (the safe and unsafe regions are separated). 
Then we have $\suffpre{s}{(P)} = \necpre{s}(P)$ since 
$\necpre{s}(P) \models_{\theory} \neg \necpre{u}(P)$. 
That is, necessary and sufficient conditions are the same for the validity 
of the assertion. 
Analogously, we have $\suffpre{u}{(P)} = \necpre{u}(P)$. 

The shared region characterised by $\phi_c$ indicates imprecision of 
over-approximations---which we attempt to reduce as much as possible. 
We achieve this reduction as follows: 
\begin{enumerate}
\item
Construct a revised program $P^I_{\phi_c}$ (Def. \ref{def:repis}) 
from $P$ focusing only on the shared region such that its SP is a 
valid SP for $P$ (Lemma \ref{prop:under-approx}). 
\item
Shrink either of the regions via iterative strengthening of the initial 
clauses of the program from where necessary preconditions are derived. 
\end{enumerate}
For this we utilize well-known CHC transformations, from the literature on 
CLP and Horn clause verification, as outlined below. 
\begin{proposition}
\label{prop:under-approx}
\rm
Let $P$ be a program, $\phi$ a constraint and $P^I_{\phi}$ as defined in 
Def.~\ref{def:repis}. 
Let $\psi$ be any SP for the safety (unsafety) of $P^I_{\phi}$. 
Then $\psi$ is also an SP for the safety (unsafety) of $P$. 
\end{proposition} 

\paragraph{1. Partial Evaluation (PE).}
\label{pe}
The PE algorithm we use \cite{Gallaghervpt19} produces a polyvariant 
specialisation, that is, a finite number of versions of each predicate, 
which is essential for deriving disjunctive information as well as for 
refining the control-flow of the program \cite{DBLP:journals/tplp/DomenechGG19}.
The result of applying PE to the example program in Fig.~\ref{ex:precond} 
wrt.\ \unsafe~and \safe~is shown in Fig.~\ref{pe_example}. 
For details we refer to \citeN{Gallaghervpt19}.
A key point is that, owing to polyvariant specialisation, \texttt{init} 
and \texttt{wh} are split into two different versions, leading to more 
precise preconditions as in Eq.~(\ref{eq:precond_pe}) 
(using Def.~\ref{def:necsafe}). 

\begin{figure}[t]
\centerline{
  \begin{tabular}{ll}
    \pcode[\small]{
$\mathtt{error \leftarrow B<0,A\leq0,wh\_2(A,B)}$. \\
$\mathtt{wh\_2(A,B) \leftarrow B<0,A \leq 0,init\_2(A,B)}$. \\
$\mathtt{wh\_2(A,B) \leftarrow B<0,A=0,C=1,}$ \\
~~~~~~~~~~~~~~~~~~~$\mathtt{B-D= -1,wh\_1(C,D)}$. \\
$\mathtt{wh\_1(A,B) \leftarrow A \geq 1,init\_1(A,B)}$. \\
$\mathtt{wh\_1(A,B) \leftarrow A \geq 1,A-C= -1,}$ \\
~~~~~~~~~~~~~~~~~~~$\mathtt{B-D= -1,wh\_1(C,D)}$. \\
$\mathtt{init\_1(A,B) \leftarrow A \ge 1}.$ \\ 
$\mathtt{init\_2(A,B) \leftarrow A \le 0, B < 0}.$
    } &
    \pcode[\small]{
$\mathtt{exit0 \leftarrow B \geq 0,A\leq0,wh\_2(A,B)}$. \\
$\mathtt{wh\_2(A,B) \leftarrow B \geq 0,A \leq 0,init\_2(A,B)}$. \\
$\mathtt{wh\_2(A,B) \leftarrow B \geq 0,A=0,C=1,}$ \\
~~~~~~~~~~~~~~~~~~~$\mathtt{B-D= -1,wh\_1(C,D)}$. \\
$\mathtt{wh\_1(A,B) \leftarrow A \geq 1,B \geq 0,init\_1(A,B)}$. \\
$\mathtt{wh\_1(A,B) \leftarrow B \geq 0,A \geq 1,A-C= -1,}$ \\
~~~~~~~~~~~~~~~~~~~$\mathtt{B-D= -1,wh\_1(C,D)}$. \\
$\mathtt{init\_1(A,B) \leftarrow A \geq 1,B \geq 0}.$ \\ 
$\mathtt{init\_2(A,B) \leftarrow B \geq 0,A \leq 0}.$
} 
  \end{tabular}
}
\caption{Partially evaluated programs: wrt.\ \texttt{error} (left) 
and wrt.\ \texttt{exit0} (right)\label{pe_example}}
\end{figure}

\begin{equation}
\label{eq:precond_pe}
\begin{array}{lcl}
 \phi_u= \mathtt{(B<0 \wedge A\leq 0) \vee 
	A\geq 1 \equiv B<0 \vee A\geq 1} \\
    \phi_s = \mathtt{(B\ge0 \wedge A\leq 0) \vee 
	(A\geq 1 \wedge B\geq 0) \equiv B\geq 0}
\end{array}
\end{equation}

\paragraph{2. Constraint Specialisation (CS).}
\label{cs}

\begin{figure}[t]
\centerline{
  \begin{tabular}{ll}
& \\
    \pcode[\small]{ 
$\mathtt{wh\_1(A,B) \leftarrow \underline{A>B}, A \ge 1, init\_1(A,B)}$. \\ 
$\mathtt{wh\_1(A,B) \leftarrow \underline{A>B}, A \ge 1, A-C= -1, }$  \\
~~~~~~~~~~~~~~~~~~$\mathtt{ B-D= -1, wh\_1(C,D)}$.  \\
$\mathtt{init\_1(A,B) \leftarrow \underline{A>B},  A \ge 1}.$ \\
 $\mathtt{init\_2(A,B) \leftarrow B < 0,A \leq 0}.$ 
    } &
    \pcode[\small]{
$\mathtt{wh\_1(A,B) \leftarrow A \ge 1,\underline{B \ge A},init\_1(A,B)}$. \\ 
$\mathtt{wh\_1(A,B) \leftarrow \underline{B \ge A},A \ge 1,A-C= -1,}$ \\
~~~~~~~~~~~~~~~~~~$\mathtt{B-D= -1,wh\_1(C,D)}$. \\
$\mathtt{init\_1(A,B) \leftarrow A \ge 1,\underline{B \ge A}}.$ \\
 $\mathtt{init\_2(A,B) \leftarrow B \geq 0,A \leq 0}.$
} 
  \end{tabular}
}
\caption{Constraint specialised programs: wrt.\ \texttt{error} (left) 
and wrt.\ \texttt{exit0} (right)\label{cs_example}}
\end{figure}

CS \cite{DBLP:journals/scp/KafleG17} of program $P$ wrt.\ goal $A$ 
specialises each constraint $\phi$ in a clause of $P$ to a constraint 
$\phi \wedge \psi$ while preserving the derivation of $A$. 
Fig.~\ref{cs_example} shows application of CS
to Fig.~\ref{pe_example} (left) wrt.\ \unsafe, and to Fig.~\ref{pe_example} 
(right) wrt.\ \safe. 
Note that only clauses that got specialised are shown, 
and the newly derived constraints are underlined for readability. 
The underlined constraints in Fig.~\ref{cs_example} (left) are obtained 
by recursively propagating $\mathtt{B <0, A \leq 0}$ top-down from the goal 
$\mathtt{error}$ and $\mathtt{A \geq 1}$ bottom-up from the initial clause 
using program transformation and abstract interpretation over the domain of 
convex polyhedra. 
In more detail, we first compute the query-answer transformed version
\cite{Codish_meta-circular} of the program in Fig.~\ref{pe_example} (left) 
wrt.\ the goal $\mathtt{error}$
(thus simulating the top-down computation), then apply abstract interpretation. 
An excerpt of the query-answer transformed program 
(just enough to show the provenience of the constraint $\mathtt{A>B}$)
is shown in Fig.~\ref{qa_example}. 

\begin{figure}[t]
\centerline{
    \pcode[\small]{
    $\mathtt{wh\_2\_q(A,B) \leftarrow B<0, A \leq 0.}$ \\
    $\mathtt{wh\_1\_q(A,B) \leftarrow D<0, C=0, A=1, B=D+1,wh\_2\_q(C,D).}$ \\
    $\mathtt{wh\_1\_q(A,B) \leftarrow C \geq 1, A=C+1, B=D+1,wh\_1\_q(C,D).}$
    }
}
\caption{Excerpt of query clauses corresponding to the program in 
Fig.~\ref{pe_example} (left) starting from the goal (query) $\unsafe$; 
the suffix q denotes a query predicate. \label{qa_example}}
\end{figure}
Since $\mathtt{wh\_1\_q}$ is the only recursive predicate, 
the rest can be unfolded away, leaving two clauses: 
$\mathtt{wh\_1\_q(A,B) \leftarrow A=1, B \leq 0}$ and 
$\mathtt{wh\_1\_q(C+1,D+1) \leftarrow wh\_1\_q(C,D), C \geq 1}$. 
Abstract interpretation using the polyhedral domain derives 
$\mathtt{A\geq 1,A>B}$ as invariant for $\mathtt{wh\_1\_q(A,B)}$ 
since we have the constraint $\mathtt{A=1, B \leq 0}$ in the base case, 
while $\mathtt{A}$ and $\mathtt{B}$ are incremented in lockstep 
in the recursive case. 

The computed invariant $\mathtt{A>B}$ for $\texttt{wh\_1\_q(A,B)}$ in 
derivations of $\unsafe$ is conjoined to each call of $\texttt{wh\_1}$, 
since the invariant holds in each such call. 
The underlined constraint $\mathtt{B\geq A}$ in Fig.~\ref{cs_example} (right) 
is obtained in similar way. 
Using these specialised programs, we derive the necessary preconditions:
\begin{equation}
\label{eq:precond_cs}
\begin{array}{lcl}
   \phi_u = \mathtt{(B<0 \wedge A\leq 0) \vee (A\geq 1 \wedge A> B)}
\\ \phi_s = \mathtt{(B\ge0 \wedge A\leq 0) \vee (A\geq 1 \wedge B\geq A)}
\end{array}
\end{equation}

\paragraph{3. Trace Elimination (TE).}
\label{te}
TE refines a program $P$ by eliminating a set of AND-trees from $P$ 
while preserving the rest of its AND-trees. 
While the elimination of infeasible trees does not have any effect 
on preconditions, extra care must be taken while eliminating feasible ones. 
Lemma~\ref{lemma:feasible} allows us to derive a safe precondition in this case.

\begin{definition}[$\theta_t$]
\label{def:trace_cs_init}
\rm
Let $P$ be a program and $t$ a \emph{feasible} AND-tree 
derived from $P$ for $\safe$ or $\unsafe$. 
Let $p^I(\tuplevar{x})$ be the atom label of an initial node of $t$. 
Then $\theta_t = \constr(t)\vert_\tuplevar{x}$ is a necessary 
condition for $t$ to be feasible. 
\end{definition}

\begin{lemma}[Adapted from \citeN{kafle-iclp18} for AND-tree of $\safe$]
\label{lemma:feasible}
\rm
Let $P'$ be the result of eliminating a \emph{feasible} AND-tree $t$ 
for $\mathtt{\safe}$ (resp.\ $\mathtt{\unsafe})$ from $P$. 
Then $\necpre{s}(P) =\necpre{s}(P') \vee \theta_t$ 
(resp.\ $\necpre{u}(P) =\necpre{u}(P') \vee \theta_t$), where 
$\theta_t$ is a constraint extracted from $t$ (Def.~\ref{def:trace_cs_init}).
\end{lemma}
Observe that the elimination of feasible traces acts as program decomposition. 
Transformations such as PE, CS and TE 
(when used to remove \emph{infeasible} trees)
not only preserve the goal but also the initial clauses. 
This allows us to construct a sequence of clauses $P_0,P_1,\ldots,P_m$ 
where $P = P_0$ and each element of the sequence is more specialised than its 
predecessor wrt.\ derivations of \safe~($\unsafe$).  
As a consequence, the NPs are more precise.
We write $P \Longrightarrow_A P'$ when $P'$ is a goal-preserving transformation
of $P$ wrt.\ an atom $A$, that is, $P \models A ~\text{iff}~ P' \models A$. 
TE (eliminating \emph{feasible} trees) is a little different, in that it 
does not preserve the goal. 
We abuse the notation and write $P \Longrightarrow_{t_A} P'$ for transformation
of $P$ by eliminating a feasible tree rooted at $A$, yielding $P'$. 
Lemma~\ref{lemma:feasible} ensures soundness of preconditions in this case. 

Let us now wrap these transformations and their combinations.
Let \tr\ and \trseq\ be any functions satisfying the following:

\begin{align*}
\trA \tuple{P,\phi} &= \left\{
\begin{array}{ll}
   \tuple{P', \phi} & \text{where~} P \Longrightarrow_A P' \\
   \textit{or} \\
   \tuple{P', \phi'} & \text{where~} P \Longrightarrow_{t_A} P' ~\text{and}~ 
        \phi'=\phi \vee \theta_{t_A} \text{(Def.\ \ref{def:trace_cs_init})}
\end{array} \right.
\\[1ex]
\trseqA \tuple{P, \phi} &= \trA^n \tuple{P, \phi} \text{~for~} n\geq 1 
\end{align*}

\noindent
where $f^n = f^{n-1} \circ f$, with $f^1 = f$.
\trseq\ allows us to combine the above transformations in any order and 
Proposition~\ref{prop:sequence} allows us to derive more precise 
preconditions. 

\begin{proposition}\label{prop:sequence}
\rm
Let $P$ be a program, $\tuple{P_s,\phi_s} = \trseqA[\safe] \tuple{P,\falseit}$.
Then $ \models_{\theory} 
(\necpre{s}{(P_s)} \vee \phi_s) \rightarrow \necpre{s}{(P)}$.
Similarly, if $\tuple{P_u, \phi_u} = \trseqA[\unsafe] \tuple{P,\falseit}$, then
$\models_{\theory} (\necpre{u}{(P_u)} \vee \phi_u) \rightarrow \necpre{u}{(P)}$.
\vspace{-3mm}
\end{proposition}

\section{An algorithm for precondition inference}
\label{sec:algorithm}

\begin{algorithm}[t]
\caption{Inferring sufficient preconditions}
\label{alg:precond}
\begin{algorithmic}[1]
\State \textbf{Input}: 
	Program $P$ with clauses for \safe, \unsafe\ and \init; 
	and trans.\ seq.\ \trseq.
\State \textbf{Output:} Pair of SPs for the safety and unsafety of $P$ 
	(wrt.\ \safe\ and \unsafe).
\State \textbf{Initialisation:} $\psi_s \gets \falseit$; 
	$\psi_u \gets \falseit$; $P_s \gets P$; $P_u \gets P$; \\
	~~~~~~~~~~~~~~~~~~~~~$\phi_{old} \gets \necpre{s}{(P)}$ 
	(Definition \ref{def:nec-suff-precond}); $\mathsf{itr} \gets 0$;
\While{true}
\State 
\hspace*{-8pt} \begin{tabular}{|l||l|l} \cline{1-2}
 $\tuple{P_s, \theta_s} \gets \trseqA[\safe]\tuple{P_s,\falseit}$ ~~~&$\tuple{P_u, \theta_u} \gets \trseqA[\unsafe]\tuple{P_u,\falseit}$ &~~~~~~~~~~~~~~~~~~~~~~~~~~~~~~~~~~~~\\ 
 $\phi_s \gets \necpre{s}{(P_s)} \vee \theta_s$ ~~~~~~~~~~~~~~~~~~~&$\phi_u \gets \necpre{u}{(P_u)} \vee \theta_u$ &~~ \\ 
\cline{1-2}
\end{tabular}
\State $\phi_{new}\gets \phi_s \wedge \phi_u$
\If{$\phi_{new}\equiv \falseit$} 
	\Comment \emph{separating} condition reached 
\State
\hspace*{-8pt} \begin{tabular}{|l||l|l}
\cline{1-2}
 $\psi_s \gets \psi_s \vee \phi_s$ ~~~& $\psi_u \gets \psi_u \vee \phi_u$ &~~~~~~~~~~~~~~~~~~~~~~~~~~~~~~~~~~~~~~~~~~~~~~~~~~~~~~~~~~~~~~~~~~~~~~~ \\ 
\cline{1-2}
 \end{tabular}
\State \Return $\tuple{\psi_s,{\psi_u}}$
\EndIf
\If{$\phi_{old} \models_{\theory} \phi_{new}$} 
	\Comment approximation was not strengthened
\State 
\hspace*{-8pt} \begin{tabular}{|l||l|l}
\cline{1-2}
$\psi_s \gets \psi_s \vee (\phi_s \wedge \neg \phi_u )$ ~~~& $\psi_u \gets \psi_u \vee (\phi_u \wedge \neg \phi_s)$ &~~~~~~~~~~~~~~~~~~~~~~~~~~~~~~~~~~~~~~~~~~~~~~~~~~~~~~~~~~~~~~~~~~~~~~~~~~~~~~~~~\\ 
\cline{1-2}
\end{tabular}
\State \Return $\tuple{\psi_s,{\psi_u}}$
\EndIf

\Comment refine programs by constraining initial clauses with $\phi_{new}$ 
	(Def.\ \ref{def:repis})
\State 
\hspace*{-8pt} \begin{tabular}{|l||l|l}
\cline{1-2}
$\psi_s \gets \psi_s \vee (\phi_s \wedge \neg \phi_u )$ ~~~& $\psi_u \gets \psi_u \vee (\phi_u \wedge \neg \phi_s)$ &~~~~~~~~~~~~~~~~~~~~~~~~~~~~~~~~~~~~~~~~~~~~~~~~~~~~~~~~~~~~~~~~~~~~~~~\\
$P_s \gets {P_{s}}_{\phi_{new}}^I$~~~~& $P_u \gets {P_{u}}_{\phi_{new}}^I$ &~~~~~~~~~~~~~~~~~~~~~~~~~~~~~~~~~~~~~~~~~~~~~~~~~~~~~~~~~~~~~~~~~~~~~~~\\ 
\cline{1-2}
\end{tabular}
\State $\phi_{old} \gets \phi_{new}; ~~~~~~\mathsf{itr}++$;
\EndWhile
\end{algorithmic}
\end{algorithm}

We now give an algorithm for computing SPs for safety and unsafety as 
Algorithm~\ref{alg:precond} based on the transformations described previously.
Input is a set of CHCs (involving clauses for \safe, \unsafe\ and \init) 
and a sequence of transformations \trseq.
Output is a pair of SPs for safety and unsafety. 
The SPs $\psi_s$ and $\psi_u$ are initialised to $\falseit$. 
The algorithm aims to weaken these SPs as far as it can. 
$\phi_{old}$ keeps track of the set of initial states that are 
yet to be proven safe or unsafe.
$P_s$ and $P_u$ respectively keep track of the transformations of 
$P$ with respect to \safe\ and \unsafe. 

In the algorithm, the following operations are carried out in an 
iterative manner and possibly in parallel (within the \emph{while} loop). 
The instructions on two sides of the boxes can be executed in parallel. 
One or more of the transformation of $P_s$ and $P_u$ with respect to \safe\ 
and \unsafe, respectively, are carried out and the NPs are extracted from 
the resulting programs (\emph{line 6}). 
The algorithm terminates and returns an SP if the conjunction of these NPs
is unsatisfiable (\emph{line 10}, \emph{separating}) or it is not stronger 
(wrt.\ $\models_{\theory}$) than $\phi_{old}$ (\emph{line 13}). 
Otherwise, the algorithm iterates with revised programs obtained by 
constraining their initial clauses with the conjunction $\phi_{new}$ 
(\emph{line 14}). 
For this, $\phi_{new}$ needs to be converted to DNF that may blow up 
the number of resulting \emph{initial clauses}. 
In our experiments, the largest size of DNF discovered was 11. 
Even if all transformations in the algorithm terminate, 
it may still not terminate since $\phi_{new}$ can infinitely be decreased.
But it makes a progress, that is, it explores a strictly smaller set of 
initial states in each iteration that have not yet been known safe or unsafe. 
This is formalised in Proposition \ref{prop:progress}. 
Observe that each iteration computes valid SPs for the safety and 
unsafety of the original program (e.g., \emph{line 14}) and combines 
them disjunctively with the previous SPs. 
Proposition \ref{prop:disj_precond} ensures that the combination yields 
valid SPs for the original program. 

\begin{proposition}[Progress and Termination of Algorithm \ref{alg:precond}]
\label{prop:progress}
\rm
Algorithm \ref{alg:precond} either progresses or terminates. 
\end{proposition}
\begin{proposition}[Composing Preconditions]
\label{prop:disj_precond}
\rm
Let $\Phi$ be a set of formulas such that each $\phi \in \Phi$ is an 
SP for (un)safety of $P$.
Then $\bigvee \Phi$ is also an SP for (un)safety of $P$. 
\end{proposition}

\noindent
Proposition \ref{prop:sequence} ensures the correctness of the 
transformations sequence, Proposition \ref{prop:under-approx} ensures 
that the precondition of $P^I_\phi$ is also that of $P$, and 
Proposition \ref{prop:disj_precond} allows us to combine the 
preconditions derived in the separate iterations. 
Together they ensure the soundness of Algorithm \ref{alg:precond}: 

\begin{theorem}[Soundness of Algorithm \ref{alg:precond}]
\label{prop:soundness}
\rm
Let $P$ be a program annotated with the predicates 
\safe\ (set of safe terminating states), 
\unsafe\ (set of unsafe terminating states)
and \init\ (set of initial states). 
If Algorithm \ref{alg:precond} returns a tuple $\tuple{S,U}$, 
then $S$ and $U$ are the SPs for safety and unsafety of $P$, 
respectively, with respect to the predicates \safe, \unsafe\ and \init.
\end{theorem}

\noindent
\textbf{Impact of transformation sequence on preconditions}. 
Let us apply the algorithm to the program $P$ in Fig.~\ref{ex:precond}.  
Initially, $\phi_{old}=\trueit$, the initial state of $P$. 
First, we choose to apply PE wrt.\ \unsafe\ and wrt.\ \safe, 
obtaining the set of CHCs shown in Fig.~\ref{pe_example}. 
The corresponding NPs are given in Eq.~(\ref{eq:precond_pe}), from which 
$\phi_{new}= \phi_s \wedge \phi_u \equiv \mathtt{B\geq 0 \wedge A \ge 1}$. 
Since $\phi_{new}$ is \emph{satisfiable}, the preconditions are not separating.
Neither is condition $\phi_{old} \models_{\theory} \phi_{new}$ satisfied, 
so the algorithm progresses to refinement (\emph{line 14-16}). 
At this point, we compute SPs for both the safety and unsafety as below. 
\begin{flalign*}
   \psi_u &= 
    \mathtt{\falseit \vee ((B<0 \vee A\geq 1) \wedge \neg (B\geq 0)) \equiv B<0}
\\ \psi_s &= 
    \mathtt{\falseit \vee (B\geq 0 \wedge \neg (B<0 \vee A\geq 1)) \equiv B\ge 0 \wedge A\leq 0} 
\end{flalign*}
As the next step, we refine $P$ to $P^I_{\phi_{new}}$, in which the clause 
for $\mathtt{init\_1}$ in Fig.~\ref{pe_example} gets strengthened to 
$\mathtt{init\_1(A,B) \leftarrow A \ge 1, B \geq 0}$. 
The clause for $\mathtt{init\_2}$ gets eliminated due to an unsatisfiable 
constraint in its body. 
The refined programs are trivial and are omitted.
In the next iteration, we apply CS with respect to \unsafe\ and \safe, 
respectively, obtaining the clauses shown in Fig.~\ref{cs_after_ris}. 
Note that the clause $\mathtt{wh\_2(A,B) \leftarrow B<0,A \leq 0,init\_2(A,B)}$
is removed since it is no longer feasible without the initial clause 
for $\mathtt{init\_2}$.
\begin{figure}[t]
\centerline{
  \begin{tabular}{ll}
    \pcode[\small]{
$\mathtt{\unsafe \leftarrow A<0, B=0, wh\_2(B,A) }$. \\ 
$\mathtt{wh\_2(A,B) \leftarrow B<0, A=0, C=1,}$ \\
~~~~~~~~~~~~~~~~$\mathtt{B-D= -1, wh\_1(C,D)}$. \\ 
$\mathtt{wh\_1(A,B) \leftarrow A>B, B\geq 0, A\geq 1, init\_1(A,B)}$. \\ 
$\mathtt{wh\_1(A,B) \leftarrow A>B, A\geq 1, A-C= -1, }$ \\
~~~~~~~~~~~~~~~~$\mathtt{B-D= -1, wh\_1(C,D)}$. \\ 
$\mathtt{init\_1(A,B)\leftarrow A>B, B\geq 0}$.  
    } &
    \pcode[\small]{
$\mathtt{\safe \leftarrow A\geq0, B=0, wh\_2(B,A)}$. \\ 
$\mathtt{wh\_2(A,B) \leftarrow B\geq0, A=0, C=1, }$ \\ 
~~~~~~~~~~~~~~~~$\mathtt{B-D= -1,wh\_1(C,D)}$. \\ 
$\mathtt{wh\_1(A,B) \leftarrow B\geq A, A\geq 1, init\_1(A,B)}$. \\ 
$\mathtt{wh\_1(A,B) \leftarrow B \geq A, A-C= -1, }$ \\  
~~~~~~~~~$\mathtt{A \geq 1,B-D= -1, wh\_1(C,D)}$. \\ 
$\mathtt{init\_1(A,B) \leftarrow B \geq A, A \geq 1}$.
    } 
  \end{tabular}
}
\vspace{-1mm}
\caption{Constraint specialised programs: 
wrt.\ $\unsafe$ (left) and wrt.\ $\safe$ (right)\label{cs_after_ris}}
\end{figure}

From these we derive: 
$ \phi_u = \mathtt{(B \ge 0 \wedge A > B)}$ and
$ \phi_s = \mathtt{(B\ge A \wedge A\geq 1)}.$ 
Since we now have $\phi_u \wedge \phi_s \equiv \falseit$, 
the preconditions are separating and the algorithm terminates. 
The final SPs are derived as the disjunction of SPs over the iterations, 
as follows: 
$\psi_u = \mathtt{B<0 \vee (B \ge 0 \wedge A > B)}$ and 
$\psi_s = \mathtt{(B\ge 0 \wedge A \leq 0) \vee (B\ge A \wedge A\geq 1)}$.
 
If instead we apply CS$\circ$PE to the original program at \emph{line 6} 
(rather than applying single transformation in each iteration), we obtain
\emph{separating} preconditions in a single iteration as shown in 
Eq.~(\ref{eq:precond_cs}), 
where $\phi_s \wedge \phi_u$ is \emph{unsatisfiable}. 
This suggests that a well chosen transformation sequence may reduce 
refinement iterations and also avoid the costly DNF conversion needed 
at \emph{line 14}.
Based on this, we fix \trseq\ to be TE$\circ$CS$\circ$PE during the 
experiments.
Our experience shows that CS is most effective when performed after PE 
which not only performs control-flow refinement of the program but also 
brings polyvariant specialisation. 
TE on the other hand helps decompose problem in addition to 
splitting predicates. 
But since it is an expensive operation, we apply it at last. 

\paragraph{Non-termination. \label{non-termination}}
The SPs derived by our method may include non-terminating inputs, 
that neither lead to safe nor unsafe.
\citeN{PopeeaC13-dualanalysis} treat such inputs as 
\emph{unsafe} whereas \citeN{SeghirS14-precond} 
ignore them, as do we. 
However, the modelling of safe and unsafe terminating states and their 
over-approximations allow us to reason about a limited form of 
non-termination as suggested by \citeN{PopeeaC13-dualanalysis}:
Any input state that is neither in the over-approximation of 
safe nor unsafe leads to non-termination 
assuming we model all terminating (un)safe states. 

\begin{wrapfigure}[8]{r}{0.39\textwidth}
\centering{
 \vspace*{-2ex}
    \pcode[\small]{
   \textbf{void} main(\textbf{int}~ a) \{ \\
   ~~\textbf{while} ($a \geq 0 $)  \{\\
   ~~~~\textbf{if} ($a \leq 9$) 
   ~~{a++;} \\
   ~~~~\textbf{else if} ($a==10$) \\
   ~~~~~~{a = 5;} \\
   ~~~~\textbf{else}~\textbf{return}; \\
   ~~\} \\
   ~~\textbf{assert} ($\id{false}$); \\
   \}          
      }    
}
  \caption{Non-termination\label{ex:non-termination}}
\end{wrapfigure}
We demonstrate this with Fig.~\ref{ex:non-termination}. 
The program does not terminate if $a \in [ 0,10 ]$. 
We derive $\phi_u= \mathtt{a} < 0$ and
$\phi_s= \mathtt{a} \geq 11$ as NPs. 
Thus the condition satisfying $\neg (\phi_u \vee \phi_s)$, that is, 
$\mathtt{a \in [ 0,10 ]}$ is a sufficient precondition for non-termination 
(which happens to be the exact condition in this case). 
It is obtained as a byproduct of our method; 
we leave the primary analysis of non-termination for future work.

\section{Experimental evaluation}
\label{sec:experiments}
Since we model both the safe and unsafe program states and successively 
refine them to be able to detect separating or more precise preconditions 
for the safety and unsafety of programs, the experiments were designed 
to \emph{better} answer the following questions.
\begin{itemize}
\item 
Q1. Does the algorithm allow us to derive separating preconditions 
in practice?
\item 
Q2. Does refinement allow us to derive more non-trivial or 
separating preconditions?
\item 
Q3. How does our approach compare to that of state-of-the-art tools for 
precondition inference in terms of the quality of the preconditions and 
performance?
\end{itemize}

\paragraph{Experimental Setup.} 
We implemented Algorithm \ref{alg:precond} (a sequential version) 
in \pihorn.\footnote{
  ``Precondition Inferrer for Horn clauses'', 
  available at \url{https://github.com/bishoksan/PI-Horn}.}
The implementation applies the sequence TE$\circ$CS$\circ$PE.
The tool is written in Ciao Prolog \cite{Ciao}, using 
PPL~\cite{BagnaraHZ08SCP} and Yices2~\cite{Dutertre:cav2014}. 
Input is a set of CHCs, with \safe, \unsafe\ and \init\ as 
distinguished predicates. 
\pihorn\ outputs a pair of SPs for safety and unsafety and are classified as: 
(i) \emph{optimal}: the precondition is both necessary and sufficient (exact); 
(ii) \emph{non-trivial}: the precondition is different from $\falseit$ 
(but not \emph{optimal}) and 
(iii) \emph{trivial}: the precondition is $\falseit$. 

Experiments were conducted on a MacBook Pro, 
running OSX 10.11 with 16GB memory and 2.7 GHz Intel Core i5 processor. 
We tested our approach with 261 integer programs (available from 
\url{https://github.com/bishoksan/PI-Horn/tree/master/benchmarks}) 
sourced as follows: 
(i) 150 integer programs from the \emph{loop} (69) and \emph{recursive} 
(81) subcategories of the \emph{Integers and Control Flow}
category of SV-COMP \cite{svcomp21}; 
(ii) 83 programs from the DAGGER \cite{DBLP:conf/tacas/GulavaniCNR08} and 
TRACER tools \cite{DBLP:conf/cav/JaffarMNS12} and 
(iii) 28 programs from the literature on precondition inference and 
backwards analysis 
\cite{DBLP:conf/sas/BakhirkinBP14,DBLP:journals/entcs/Mine12,DBLP:conf/vmcai/Moy08,DBLP:conf/sas/BakhirkinM17,DBLP:conf/rp/CassezJL17}. 
We are unable to include some benchmarks used by \citeN{kafle-iclp18} 
owing to unavailability of their C sources which are needed to model the 
exit states. 
Benchmark set (i) was designed for verification competitions, 
(ii) and (iii) to demonstrate particular tools and techniques. 
We adapt these C programs for precondition inference as follows. 
They are translated to CHCs of the required form based on 
specialisation approach of \citeN{DBLP:journals/scp/AngelisFPP17} 
using VeriMap \cite{verimap}. 
We then replace the generated 
$\init(\tuplevar{x}) \leftarrow \phi(\tuplevar{x})$ clause by 
$\init(\tuplevar{x}) \leftarrow \trueit$.
This allows analyses to infer preconditions in terms of $\tuplevar{x}$, 
starting from an unrestricted set of \emph{initial} clauses. 

\paragraph{Results and Discussion.}
Table \ref{tbl:exp-results} shows the results. 
The columns 2-7 show results for \pihorn\ and the 
last column for \wprahft~\cite{kafle-iclp18}. 
The first column \textsf{iter} indicates the number of 
refinement iterations for both. 
The columns show
\textsf{opt} (\# programs with separating preconditions), 
\textsf{ntS} (\textsf{Sw}) (\# programs with non-trivial SPs for 
safety excluding separating cases, and, in parentheses, 
the difference with trivial SPs for unsafety),
\textsf{ntU} (\textsf{Uw}) (same, for unsafety), 
\textsf{ntSU} (\# programs with non-trivial SPs for both safety and unsafety), 
\textsf{tSU} (\# programs with trivial SPs for both safety and unsafety), 
\textsf{total}/\textsf{iter} (\# programs with non-trivial 
(either for safety or unsafety) plus separating SP per iteration), 
\wprahft~\textsf{total}/\textsf{iter} (\# programs with non-trivial SP 
per iteration for \wprahft). 
For example, the entry 9 (7) in column 3 indicates that there were 9
non-trivial SPs for safety, of which 7 had trivial SPs for unsafety. 
In other words, the number in the parentheses counts the pairs of the 
form $\langle \textsf{ntS},\textsf{tU} \rangle$ 
where \textsf{tU} means trivial precondition for unsafety. 
Each row corresponds to an iteration and contains the number of instances 
solved in that iteration, excluding those solved in the previous.

\begin{table}[t]
\newcommand{\ph}{\hphantom{1}}
\begin{center}
    \begin{tabular}{|r|r|r|r|r|r|r|r|r|}
    \cline{1-9}
   \centering{\textsf{iter}} & \textsf{opt} & {\centering \textsf{ntS}  (\textsf{Sw})} & {\centering \textsf{ntU} (\textsf{Uw})} & {\centering \textsf{ntSU}} & {\centering \textsf{tSU} }&&   \multicolumn{1}{p{2cm}|}{\centering \pihorn  \\ \textsf{total}/\textsf{iter}} & \multicolumn{1}{p{2cm}|}{\centering \wprahft  \\ \textsf{total}/\textsf{iter}} \\ 
\cline{1-9} 
    0  & 58  & 0 \ph(0)   & 0 (0)  & 0  & 0   && 58  & 197 \\ \cline{1-9}
    1  & 87  & 9 \ph(7)   & 5 (3)  & 2  & 20  && 99  &  20 \\ \cline{1-9}
    2  & 21  & 20 (15)    & 7 (2)  & 5  & 0   && 43  &   0 \\ \cline{1-9}
    3  & 5   & 6 \ph(3)   & 3 (0)  & 3  & 0   && 11  &   0 \\ \cline{1-9}
    4  & 2   & 3 \ph(1)   & 2 (0)  & 2  & 0   && 5   &   0 \\ \cline{1-9}
    5  & 2   & 0 \ph(0)   & 0 (0)  & 0  & 0   && 2   &   1 \\ \cline{1-9}
    6  & 1   & 0 \ph(0)   & 0 (0)  & 0  & 0   && 1   &   0 \\ \cline{1-9}
    \#total & 176  & 38 (26)  & 17 (5)  & 12  & 20  && 219  & 218 \\ \cline{1-9}
    \end{tabular}
\end{center}
\vspace{-2mm}
\caption{Experimental results on 261 programs, 
with a timeout of 300 seconds\label{tbl:exp-results}}
\end{table}

The results answer Q1 and Q2 positively. 
\pihorn~infers non-trivial preconditions 
for 83\% and optimal ones for 67\% of the programs.  
Interestingly, it infers optimal preconditions for 58 programs owing to 
specialisation transformations alone (see row \#1, \textsf{iter} 0), 
whereas it infers non-trivial preconditions for 99 programs 
(of which 87 are optimal) after the first refinement. 
More non-trivial preconditions are derived when refinement progresses. 
This indicates that both the preprocessing and refinement significantly 
increase the number of optimal (non-trivial) cases. 
However, for 63 out of 261 programs, refinement did not progress 
towards optimality (that is, it did not further shrink the approximations 
of safe and unsafe states). 
We also observe that it timed out on 9\% and failed to infer any 
meaningful preconditions for 8\%. 

As for Q3 we could not meaningfully compare our tool against the 
work of \citeN{SeghirS14-precond}, or \citeN{DBLP:conf/sas/BakhirkinBP14}, 
in the first case because of tool issues 
(discovered together with the authors), and a lack of automation 
(confirmed by the authors via email) for the second. 
We do compare with \wprahft~\cite{kafle-iclp18}, but note that, 
while some of the components of the tools are identical, 
the results are not directly comparable. 
The success of \pihorn\ depends on its ability to refine both the 
approximations simultaneously unlike \wprahft. 
For example, we might obtain a tight bound $\phi$ for safe states 
but if the approximation of unsafe states is $\psi$ such that 
$\phi \models_{\theory}\psi, \psi \not\equiv \trueit$ then \pihorn\ returns 
trivial SP for safety whereas \wprahft\ returns non-trivial. 
On the other hand, \wprahft\ cannot detect optimality 
and does not derive preconditions for unsafety. 
The two tools provide almost the same number of
programs with non-trivial SPs (219 vs 218), 
but they differ in the quality (e.g., optimality) of preconditions. 
Since \wprahft\ cannot detect optimality, limited information about it 
can be obtained by checking 
$\suffpre{\pihorn} \models_{\theory}\suffpre{\wprahft}$ on all those 
instances that are known to be optimal, where $\suffpre{x}$ represents 
the sufficient precondition derived by the tool $x$.  
From this, we report that \wprahft\ derives optimal preconditions for 
58 programs (\# of successful checks), far less than \pihorn\ (176). 
Every refinement yields improvements for \pihorn\ but refinement 
beyond the second yields negligible improvements for \wprahft. 
This affirms that focusing attention on the intersection of 
approximations of safe and unsafe states is a good refinement strategy 
and shows the benefit of concurrently approximating these states. 
Thanks to the refined termination criteria of \pihorn\ that the average 
time in seconds per instance is 30.72 (22 timeouts), 
while for \wprahft\ it goes from 28.14 (\textsf{iter} 1, 20 timeouts) 
to 41.86 (\textsf{iter} 6, 30 timeouts). 
In summary for Q3, \pihorn\ infers \emph{better} preconditions 
than \wprahft\ and shows reasonable performance. 

\section{Related work}
\label{sec:rel}

Over-approximation techniques (forward/backward abstract interpretations or
their combination \cite{CousotCousot92-3,CousotCFL13,DBLP:conf/sas/BakhirkinM17}) 
inherently derive NPs, and complementation supplies SPs at a cost of precision 
(due to approximation of the complement). 
\citeN{DBLP:conf/lopstr/HoweKL04} use a pseudo-complemented domain (\emph{Pos}) 
domain~\cite{Marriott-Sondergaard-LOPLAS93} to infer SPs; 
\citeN{DBLP:conf/sas/BakhirkinBP14} exchange an abstract complement operation 
for abstract logical subtraction. 
Our method neither assumes an abstract domain is (pseudo-) complemented nor 
apply complementation of abstract elements during analysis. 
It applies to any abstract domain, and complementation is carried out 
externally to abstract interpretation, storing the result as a formula 
without any loss of precision.

Little work has been done that inherently computes SPs without complementation.
The notable exception is the work by \citeN{DBLP:journals/entcs/Mine12}, 
who designs all required purpose-built backward transfer functions for 
intervals, octagons and convex polyhedra domains. 
The downside is that the purpose-built operations, including widening, 
can be rather intricate and require substantial implementation effort. 
\citeN{DBLP:conf/vmcai/Moy08} employs weakest-precondition reasoning 
and forward abstract interpretation to generalise conditions at 
loop heads to infer SPs. 
The derived conditions offer limited use except for a theorem prover. 
Our method, on the other hand, uses standard techniques and 
off-the-shelf tools. 
Output from \pihorn\ can be consumed by other analysis and verification tools.

In a verification context, the dual-analysis approach of 
\citeN{PopeeaC13-dualanalysis} uses over-approximations, as we do, 
to concurrently infer NPs for safety and unsafety; from that, SPs are derived. 
No attempt is made to weaken those preconditions 
(by refining the approximations or focusing the analysis, as we do); 
we suspect such SPs are overly strong.
\citeN{DilligDLM13} use Hoare style reasoning with abduction iteratively, 
to infer loop invariants that are sufficient to show validity of assertions. 
The success of their method relies on guessing good abducibles whereas 
computing precise inductive invariants is too hard to achieve for 
realistic programs (due to undecidability).

Program transformation approaches that preserve the goal can be used to 
derive preconditions, as our approach. 
These include the forward/backward iterative specialisation by
\citeN{DBLP:journals/scp/AngelisFPP14}, for verifying program properties. 
The transformation approach uses a constraint generalisation instead of 
abstract interpretation. 
Similarly, the multivariant top-down analyzer by \citeN{PueblaH99,mcctr-fixpt}
produces polyvariant specializations, as in the classical algorithms, 
performing backwards analysis using abstract interpretation. 
These methods are complementary to ours and we leave a comparative 
study with our method for future work. 

\citeN{SeghirS14-precond} use a CEGAR approach to derive \emph{exact} 
necessary and sufficient preconditions for safety. 
Like us, they model safe and unsafe states of a program and refine 
their approximations until they are disjoint. 
Their algorithm may diverge due to (i) the lack of a suitable 
generalisation of the counterexamples (an inherent limitation of CEGAR) 
and (ii) the termination condition (disjointness) that is too hard to 
achieve for realistic programs (due to undecidability). 
\citeN{PadhiSM16} attempt to derive optimal preconditions using 
machine learning approaches. 
The success of their approach relies on learning good heuristics to 
separate good runs from bad runs. 
We, in contrast, use abstract interpretation and program transformation, 
so each step of the algorithm terminates and a sound precondition 
can be derived from the resulting programs. 
Besides, optimality is not the end goal for us and it is a 
by-product of precision refinement. 

The work of \citeN{kafle-iclp18} is orthogonal to those above, 
combining a range of established techniques such as abstract interpretation, 
CEGAR and program transformations in a profitable way. 
The iterative nature of their approach allows them to derive more 
precise preconditions for safety, however the termination criterion, 
the maximum number of iterations supplied by the user, 
is rather weak and cannot be used to optimality of the preconditions. 
The current work offers several advantages. 
We model both safe and unsafe states that enables us to detect 
optimality and also infer NP and SP for both safety and unsafety. 
In addition, it allows reasoning about a limited form of non-termination 
and provides more refined termination criteria. 
Unlike many methods in the literature 
\cite{SeghirS14-precond,DBLP:conf/sas/BakhirkinBP14}, our method can 
uniformly handle programs with procedures and recursive programs.

\section{Concluding remarks}
\label{sec:conclusion}

We have presented an iterative method for automatically deriving 
sufficient preconditions for the safety and unsafety of programs. 
It maintains over-approximations of the set of \emph{safe} and 
\emph{unsafe} initial states. 
Each iteration of the algorithm considers only states that are common to 
these approximations as they are yet to be classified as safe or unsafe. 
The method terminates when the common set of states is empty or it 
fails to shrink in successive iterations. 
In experiments, the method generated separating preconditions in $67\%$ 
of test cases and solved problems which fail to resolve using only 
approximation of unsafe states (as done in previous work).
Owing to over-approximation, the sufficient preconditions may include 
some non-terminating states, which hinders the derivation of optimal 
preconditions. 
Our method can only infer preconditions that are expressible as 
boolean combinations of (quantifier free) linear integer constraints 
and the prototype implementation mostly ignores simplification of 
preconditions, possibly leaving redundancies. 
In future work, we intend to augment our method with non-termination 
analysis, extend it to infer quantified preconditions and work on 
simplifying the preconditions.

\section*{Acknowledgements}
We thank John Gallagher and three anonymous reviewers whose suggestions
helped improve the paper.
We are also grateful for help with the use of VeriMap for C to CHC translation,
provided to us by Emanuele De Angelis.
Bishoksan Kafle has been partially funded by the Spanish Ministry of Research,
Science and Innovation, grant MICINN PID2019-108528RB-C21 \emph{ProCode} 
and Madrid P2018/TCS-4339 \emph{BLOQUES-CM}.  

\bibliographystyle{acmtrans}
\bibliography{refs}

\begin{thebibliography}{}

\bibitem[\protect\citeauthoryear{Bagnara, Hill, and Zaffanella}{Bagnara
  et~al\mbox{.}}{2008}]{BagnaraHZ08SCP}
{\sc Bagnara, R.}, {\sc Hill, P.~M.}, {\sc and} {\sc Zaffanella, E.} 2008.
\newblock The {Parma Polyhedra Library}: Toward a complete set of numerical
  abstractions for the analysis and verification of hardware and software
  systems.
\newblock {\em Sci.\ Comput.\ Program.\/}~{\em 72,\/}~1--2, 3--21.

\bibitem[\protect\citeauthoryear{Bakhirkin, Berdine, and Piterman}{Bakhirkin
  et~al\mbox{.}}{2014}]{DBLP:conf/sas/BakhirkinBP14}
{\sc Bakhirkin, A.}, {\sc Berdine, J.}, {\sc and} {\sc Piterman, N.} 2014.
\newblock Backward analysis via over-approximate abstraction and
  under-approximate subtraction.
\newblock In {\em SAS'14}. LNCS, vol. 8723. Springer, 34--50.

\bibitem[\protect\citeauthoryear{Bakhirkin and Monniaux}{Bakhirkin and
  Monniaux}{2017}]{DBLP:conf/sas/BakhirkinM17}
{\sc Bakhirkin, A.} {\sc and} {\sc Monniaux, D.} 2017.
\newblock Combining forward and backward abstract interpretation of {Horn}
  clauses.
\newblock In {\em SAS'17}. LNCS, vol. 10422. Springer, 23--45.

\bibitem[\protect\citeauthoryear{Beyer}{Beyer}{2021}]{svcomp21}
{\sc Beyer, D.} 2021.
\newblock Software verification: 10th comparative evaluation {(SV-COMP} 2021).
\newblock In {\em TACAS 2021}, {J.~F. Groote} {and} {K.~G. Larsen}, Eds. LNCS.
  Springer, 401--422.

\bibitem[\protect\citeauthoryear{Beyer, Henzinger, Majumdar, and
  Rybalchenko}{Beyer et~al\mbox{.}}{2007}]{BeyerHMR07}
{\sc Beyer, D.}, {\sc Henzinger, T.~A.}, {\sc Majumdar, R.}, {\sc and} {\sc
  Rybalchenko, A.} 2007.
\newblock Path invariants.
\newblock In {\em {PLDI}}, {J.~Ferrante} {and} {K.~S. McKinley}, Eds. {ACM},
  300--309.

\bibitem[\protect\citeauthoryear{Cassez, Jensen, and Larsen}{Cassez
  et~al\mbox{.}}{2017}]{DBLP:conf/rp/CassezJL17}
{\sc Cassez, F.}, {\sc Jensen, P.~G.}, {\sc and} {\sc Larsen, K.~G.} 2017.
\newblock Refinement of trace abstraction for real-time programs.
\newblock In {\em Reachability Problems}. LNCS, vol. 10506. Springer, 42--58.

\bibitem[\protect\citeauthoryear{Codish and S{\o}ndergaard}{Codish and
  S{\o}ndergaard}{2002}]{Codish_meta-circular}
{\sc Codish, M.} {\sc and} {\sc S{\o}ndergaard, H.} 2002.
\newblock Meta-circular abstract interpretation in {Prolog}.
\newblock In {\em The Essence of Computation}, {T.~Mogensen} {et~al\mbox{.}},
  Eds. LNCS, vol. 2566. Springer, 109--134.

\bibitem[\protect\citeauthoryear{Cousot and Cousot}{Cousot and
  Cousot}{1992}]{CousotCousot92-3}
{\sc Cousot, P.} {\sc and} {\sc Cousot, R.} 1992.
\newblock Abstract interpretation and application to logic programs.
\newblock {\em J.\ Logic Programming\/}~{\em 13,\/}~2{\&}3, 103--179.

\bibitem[\protect\citeauthoryear{Cousot, Cousot, F{\"{a}}hndrich, and
  Logozzo}{Cousot et~al\mbox{.}}{2013}]{CousotCFL13}
{\sc Cousot, P.}, {\sc Cousot, R.}, {\sc F{\"{a}}hndrich, M.}, {\sc and} {\sc
  Logozzo, F.} 2013.
\newblock Automatic inference of necessary preconditions.
\newblock In {\em VMCAI'13}. LNCS, vol. 7737. Springer, 128--148.

\bibitem[\protect\citeauthoryear{Cousot and Halbwachs}{Cousot and
  Halbwachs}{1978}]{CousotH78}
{\sc Cousot, P.} {\sc and} {\sc Halbwachs, N.} 1978.
\newblock Automatic discovery of linear restraints among variables of a
  program.
\newblock In {\em {POPL}}. {ACM} Press, 84--96.

\bibitem[\protect\citeauthoryear{{De Angelis}, Fioravanti, Pettorossi, and
  Proietti}{{De Angelis} et~al\mbox{.}}{2014}]{DBLP:journals/scp/AngelisFPP14}
{\sc {De Angelis}, E.}, {\sc Fioravanti, F.}, {\sc Pettorossi, A.}, {\sc and}
  {\sc Proietti, M.} 2014.
\newblock Program verification via iterated specialization.
\newblock {\em Sci.\ Comput.\ Program.\/}~{\em 95}, 149--175.

\bibitem[\protect\citeauthoryear{De~Angelis, Fioravanti, Pettorossi, and
  Proietti}{De~Angelis et~al\mbox{.}}{2014}]{verimap}
{\sc De~Angelis, E.}, {\sc Fioravanti, F.}, {\sc Pettorossi, A.}, {\sc and}
  {\sc Proietti, M.} 2014.
\newblock {VeriMAP}: A tool for verifying programs through transformations.
\newblock In {\em TACAS 2014}. LNCS, vol. 8413. Springer, 568--574.

\bibitem[\protect\citeauthoryear{{De Angelis}, Fioravanti, Pettorossi, and
  Proietti}{{De Angelis} et~al\mbox{.}}{2017}]{DBLP:journals/scp/AngelisFPP17}
{\sc {De Angelis}, E.}, {\sc Fioravanti, F.}, {\sc Pettorossi, A.}, {\sc and}
  {\sc Proietti, M.} 2017.
\newblock Semantics-based generation of verification conditions via program
  specialization.
\newblock {\em Sci.\ Comput.\ Program.\/}~{\em 147}, 78--108.

\bibitem[\protect\citeauthoryear{Dillig, Dillig, Li, and McMillan}{Dillig
  et~al\mbox{.}}{2013}]{DilligDLM13}
{\sc Dillig, I.}, {\sc Dillig, T.}, {\sc Li, B.}, {\sc and} {\sc McMillan,
  K.~L.} 2013.
\newblock Inductive invariant generation via abductive inference.
\newblock In {\em OOPSLA 2013}. ACM, 443--456.

\bibitem[\protect\citeauthoryear{Dom{\'{e}}nech, Gallagher, and
  Genaim}{Dom{\'{e}}nech et~al\mbox{.}}{2019}]{DBLP:journals/tplp/DomenechGG19}
{\sc Dom{\'{e}}nech, J.~J.}, {\sc Gallagher, J.~P.}, {\sc and} {\sc Genaim, S.}
  2019.
\newblock Control-flow refinement by partial evaluation, and its application to
  termination and cost analysis.
\newblock {\em Theory Pract.\ Log.\ Program.\/}~{\em 19,\/}~5-6, 990--1005.

\bibitem[\protect\citeauthoryear{Dutertre}{Dutertre}{2014}]{Dutertre:cav2014}
{\sc Dutertre, B.} 2014.
\newblock Yices 2.2.
\newblock In {\em CAV 2014}. LNCS, vol. 8559. Springer, 737--744.

\bibitem[\protect\citeauthoryear{Gallagher}{Gallagher}{2019}]{Gallaghervpt19}
{\sc Gallagher, J.~P.} 2019.
\newblock Polyvariant program specialisation with property-based abstraction.
\newblock In {\em VPT 2019}. EPTCS, vol. 299. 34--48.

\bibitem[\protect\citeauthoryear{Gallagher and Lafave}{Gallagher and
  Lafave}{1996}]{Gallagher-Lafave-Dagstuhl}
{\sc Gallagher, J.~P.} {\sc and} {\sc Lafave, L.} 1996.
\newblock Regular approximation of computation paths in logic and functional
  languages.
\newblock In {\em Partial Evaluation}. LNCS, vol. 1110. Springer, 115--136.

\bibitem[\protect\citeauthoryear{Grebenshchikov, Lopes, Popeea, and
  Rybalchenko}{Grebenshchikov
  et~al\mbox{.}}{2012}]{DBLP:conf/pldi/GrebenshchikovLPR12}
{\sc Grebenshchikov, S.}, {\sc Lopes, N.~P.}, {\sc Popeea, C.}, {\sc and} {\sc
  Rybalchenko, A.} 2012.
\newblock Synthesizing software verifiers from proof rules.
\newblock In {\em PLDI 2012}. {ACM}, 405--416.

\bibitem[\protect\citeauthoryear{Gulavani, Chakraborty, Nori, and
  Rajamani}{Gulavani et~al\mbox{.}}{2008}]{DBLP:conf/tacas/GulavaniCNR08}
{\sc Gulavani, B.~S.}, {\sc Chakraborty, S.}, {\sc Nori, A.~V.}, {\sc and} {\sc
  Rajamani, S.~K.} 2008.
\newblock Automatically refining abstract interpretations.
\newblock In {\em TACAS 2008}. LNCS, vol. 4963. Springer, 443--458.

\bibitem[\protect\citeauthoryear{Gulwani, Jain, and Koskinen}{Gulwani
  et~al\mbox{.}}{2009}]{GulwaniJK09}
{\sc Gulwani, S.}, {\sc Jain, S.}, {\sc and} {\sc Koskinen, E.} 2009.
\newblock Control-flow refinement and progress invariants for bound analysis.
\newblock In {\em {PLDI}}. {ACM}, 375--385.

\bibitem[\protect\citeauthoryear{Gupta and Rybalchenko}{Gupta and
  Rybalchenko}{2009}]{GuptaR09}
{\sc Gupta, A.} {\sc and} {\sc Rybalchenko, A.} 2009.
\newblock Invgen: An efficient invariant generator.
\newblock In {\em {CAV}}, {A.~Bouajjani} {and} {O.~Maler}, Eds. LNCS, vol.
  5643. Springer, 634--640.

\bibitem[\protect\citeauthoryear{Gurfinkel, Kahsai, Komuravelli, and
  Navas}{Gurfinkel et~al\mbox{.}}{2015}]{DBLP:conf/cav/GurfinkelKKN15}
{\sc Gurfinkel, A.}, {\sc Kahsai, T.}, {\sc Komuravelli, A.}, {\sc and} {\sc
  Navas, J.~A.} 2015.
\newblock The {SeaHorn} verification framework.
\newblock In {\em CAV 2015}. LNCS, vol. 9206. Springer, 343--361.

\bibitem[\protect\citeauthoryear{Hermenegildo, Bueno, Carro,
  et~al\mbox{.}}{Hermenegildo et~al\mbox{.}}{2012}]{Ciao}
{\sc Hermenegildo, M.~V.}, {\sc Bueno, F.}, {\sc Carro, M.}, {\sc
  et~al\mbox{.}} 2012.
\newblock An overview of {Ciao} and its design philosophy.
\newblock {\em Theory and Practice of Logic Programming\/}~{\em 12,\/}~1-2,
  219--252.

\bibitem[\protect\citeauthoryear{Howe, King, and Lu}{Howe
  et~al\mbox{.}}{2004}]{DBLP:conf/lopstr/HoweKL04}
{\sc Howe, J.~M.}, {\sc King, A.}, {\sc and} {\sc Lu, L.} 2004.
\newblock Analysing logic programs by reasoning backwards.
\newblock In {\em Program Development in Computational Logic}. LNCS, vol. 3049.
  Springer, 152--188.

\bibitem[\protect\citeauthoryear{Jaffar, Murali, Navas, and Santosa}{Jaffar
  et~al\mbox{.}}{2012}]{DBLP:conf/cav/JaffarMNS12}
{\sc Jaffar, J.}, {\sc Murali, V.}, {\sc Navas, J.~A.}, {\sc and} {\sc Santosa,
  A.~E.} 2012.
\newblock {TRACER}: A symbolic execution tool for verification.
\newblock In {\em CAV 2012}. LNCS, vol. 7358. Springer, 758--766.

\bibitem[\protect\citeauthoryear{Jones, Gomard, and Sestoft}{Jones
  et~al\mbox{.}}{1993}]{Jones-Gomard-Sestoft}
{\sc Jones, N.}, {\sc Gomard, C.}, {\sc and} {\sc Sestoft, P.} 1993.
\newblock {\em {P}artial {E}valuation and {A}utomatic {S}oftware {G}eneration}.
\newblock Prentice Hall.

\bibitem[\protect\citeauthoryear{Kafle and Gallagher}{Kafle and
  Gallagher}{2017a}]{DBLP:journals/scp/KafleG17}
{\sc Kafle, B.} {\sc and} {\sc Gallagher, J.~P.} 2017a.
\newblock Constraint specialisation in {Horn} clause verification.
\newblock {\em Sci.\ Comput.\ Program.\/}~{\em 137}, 125--140.

\bibitem[\protect\citeauthoryear{Kafle and Gallagher}{Kafle and
  Gallagher}{2017b}]{DBLP:journals/cl/KafleG17}
{\sc Kafle, B.} {\sc and} {\sc Gallagher, J.~P.} 2017b.
\newblock Horn clause verification with convex polyhedral abstraction and tree
  automata-based refinement.
\newblock {\em Comput.\ Lang.\ Syst.\ Struct.\/}~{\em 47}, 2--18.

\bibitem[\protect\citeauthoryear{Kafle, Gallagher, Gange, et~al\mbox{.}}{Kafle
  et~al\mbox{.}}{2018}]{kafle-iclp18}
{\sc Kafle, B.}, {\sc Gallagher, J.~P.}, {\sc Gange, G.}, {\sc et~al\mbox{.}}
  2018.
\newblock An iterative approach to precondition inference using constrained
  {Horn} clauses.
\newblock {\em Theory Pract.\ Log.\ Program.\/}~{\em 18}, 553--570.

\bibitem[\protect\citeauthoryear{Karr}{Karr}{1976}]{Karr76}
{\sc Karr, M.} 1976.
\newblock Affine relationships among variables of a program.
\newblock {\em Acta Informatica\/}~{\em 6}, 133--151.

\bibitem[\protect\citeauthoryear{Marriott and S{\o}ndergaard}{Marriott and
  S{\o}ndergaard}{1993}]{Marriott-Sondergaard-LOPLAS93}
{\sc Marriott, K.} {\sc and} {\sc S{\o}ndergaard, H.} 1993.
\newblock Precise and efficient groundness analysis for logic programs.
\newblock {\em ACM Letters Program.\ Lang.\ Syst.\/}~{\em 2,\/}~1--4, 181--196.

\bibitem[\protect\citeauthoryear{Min{\'{e}}}{Min{\'{e}}}{2006}]{Mine06}
{\sc Min{\'{e}}, A.} 2006.
\newblock The octagon abstract domain.
\newblock {\em High. Order Symb. Comput.\/}~{\em 19,\/}~1, 31--100.

\bibitem[\protect\citeauthoryear{Min{\'{e}}}{Min{\'{e}}}{2012}]{DBLP:journals/entcs/Mine12}
{\sc Min{\'{e}}, A.} 2012.
\newblock Inferring sufficient conditions with backward polyhedral
  under-approximations.
\newblock {\em Electronic Notes in Theor.\ Comp.\ Sci.\/}~{\em 287}, 89--100.

\bibitem[\protect\citeauthoryear{Moy}{Moy}{2008}]{DBLP:conf/vmcai/Moy08}
{\sc Moy, Y.} 2008.
\newblock Sufficient preconditions for modular assertion checking.
\newblock In {\em VMCAI 2008}. LNCS, vol. 4905. Springer, 188--202.

\bibitem[\protect\citeauthoryear{Muthukumar and Hermenegildo}{Muthukumar and
  Hermenegildo}{1990}]{mcctr-fixpt}
{\sc Muthukumar, K.} {\sc and} {\sc Hermenegildo, M.} 1990.
\newblock {D}eriving {A} {F}ixpoint {C}omputation {A}lgorithm for {T}op-down
  {A}bstract {I}nterpretation of {L}ogic {P}rograms.
\newblock Technical Report ACT-DC-153-90, Microelectronics and Computer
  Technology Corporation (MCC), Austin, TX 78759. April.

\bibitem[\protect\citeauthoryear{Padhi, Sharma, and Millstein}{Padhi
  et~al\mbox{.}}{2016}]{PadhiSM16}
{\sc Padhi, S.}, {\sc Sharma, R.}, {\sc and} {\sc Millstein, T.~D.} 2016.
\newblock Data-driven precondition inference with learned features.
\newblock In {\em PLDI 2016}. {ACM}, 42--56.

\bibitem[\protect\citeauthoryear{Peralta, Gallagher, and Sa\u{g}lam}{Peralta
  et~al\mbox{.}}{1998}]{Peralta-Gallagher-Saglam-SAS98}
{\sc Peralta, J.~C.}, {\sc Gallagher, J.~P.}, {\sc and} {\sc Sa\u{g}lam, H.}
  1998.
\newblock Analysis of imperative programs through analysis of constraint logic
  programs.
\newblock In {\em SAS 1998}. LNCS, vol. 1503. 246--261.

\bibitem[\protect\citeauthoryear{Popeea and Chin}{Popeea and
  Chin}{2013}]{PopeeaC13-dualanalysis}
{\sc Popeea, C.} {\sc and} {\sc Chin, W.} 2013.
\newblock Dual analysis for proving safety and finding bugs.
\newblock {\em Sci.\ Comput.\ Program.\/}~{\em 78,\/}~4, 390--411.

\bibitem[\protect\citeauthoryear{Puebla and Hermenegildo}{Puebla and
  Hermenegildo}{1999}]{PueblaH99}
{\sc Puebla, G.} {\sc and} {\sc Hermenegildo, M.~V.} 1999.
\newblock Abstract multiple specialization and its application to program
  parallelization.
\newblock {\em J. Log. Program.\/}~{\em 41,\/}~2-3, 279--316.

\bibitem[\protect\citeauthoryear{Sankaranarayanan, Sipma, and
  Manna}{Sankaranarayanan et~al\mbox{.}}{2004}]{SankaranarayananSM04}
{\sc Sankaranarayanan, S.}, {\sc Sipma, H.}, {\sc and} {\sc Manna, Z.} 2004.
\newblock Non-linear loop invariant generation using {Gr{\"{o}}bner} bases.
\newblock In {\em {POPL}}. {ACM}, 318--329.

\bibitem[\protect\citeauthoryear{Seghir and Schrammel}{Seghir and
  Schrammel}{2014}]{SeghirS14-precond}
{\sc Seghir, M.~N.} {\sc and} {\sc Schrammel, P.} 2014.
\newblock Necessary and sufficient preconditions via eager abstraction.
\newblock In {\em APLAS 2014}. LNCS, vol. 8858. Springer, 236--254.

\end{thebibliography}

\end{document}